\documentclass[preprint,floatfix]{article}

\usepackage{etex}
\usepackage{ifpdf}
\usepackage{hyperref}
\usepackage{dcolumn}
\usepackage{url}
\usepackage{amsmath}
\usepackage{amscd}
\usepackage{amsfonts}
\usepackage{amssymb}
\usepackage{amsthm}
\usepackage{xfrac}
\usepackage{braket}
\usepackage{bm}   
\usepackage{bbm}
\usepackage{verbatim}
\usepackage{stmaryrd}
\usepackage{xcolor}
\usepackage{setspace}
\usepackage{tikz}
\usetikzlibrary{arrows}
\usetikzlibrary{automata}
\usetikzlibrary{positioning}
\usetikzlibrary{plotmarks}
\usepackage{pgfplots}
\pgfplotsset{compat=newest}
\usepackage{graphicx}
\usepackage{sidecap}
\graphicspath{{plots/}{pictures/}}
\usepackage[normal,tight,center]{subfigure}
\usepackage[toc,page]{appendix}
\usepackage[affil-it]{authblk}
\usepackage{floatpag}
\usepackage[numbers,sort&compress]{natbib}
\tikzstyle{vaucanson}=[
  node distance=2.5cm,
  bend angle=15,
  auto,
  every loop/.style={},
  every edge/.style={->,draw=black,line width=1.2,>=latex,shorten <=1pt,shorten >=1pt},
  every state/.style={draw=black,line width=2,font=\large},
  loop right/.style={right,out=22,in=-22,loop},
  loop above/.style={above,out=112,in=68,loop},
  loop left/.style={left,out=202,in=158,loop},
  loop below/.style={below,out=292,in=248,loop},
  loop above right/.style={right,out=67,in=23,loop},
  loop above left/.style={left,out=157,in=113,loop},
  loop below left/.style={left,out=247,in=203,loop},
  loop below right/.style={right,out=337,in=293,loop},
]

\theoremstyle{plain}    
\theoremstyle{plain}    
\theoremstyle{plain}    
\theoremstyle{plain}    
\theoremstyle{plain}    
\theoremstyle{plain}    
\theoremstyle{plain}    
\theoremstyle{plain}    
\theoremstyle{plain}    
\theoremstyle{plain}    
\theoremstyle{plain}    
\theoremstyle{plain}    
\theoremstyle{plain}    

\floatpagestyle{empty}

\begin{document}

\title{Parasite Spreading in Spatial Ecological Multiplex Networks}

\author{Massimo Stella \thanks{Corresponding author: massimo.stella@inbox.com}}
\affil{Institute for Complex Systems Simulation, University of Southampton, Southampton, UK}

\author{Cecilia S. Andreazzi \thanks{Corresponding author: candreazzi@fiocruz.br}}
\affil{Departamento de Ecologia, Universidade de S\~{a}o Paulo, S\~{a}o Paulo, SP 05508-900, Brazil}
\affil{Funda\c{c}\~{a}o Oswaldo Cruz, Rio de Janeiro, RJ, 22713-375, Brazil}

\author{Sanja Selakovic}
\affil{Faculty of Veterinary Medicine, Utrecht University, Utrecht, The Netherlands}

\author{Alireza Goudarzi}
\affil{Department of Computer Science, University of New Mexico,
Albuquerque, NM 87131-0001}

\author{Alberto Antonioni}
\affil{Faculty of Business and Economics, University of Lausanne,  1015 Lausanne, Switzerland}
\affil{Grupo Interdisciplinar de Sistemas Complejos (GISC), Departamento de Matem\'aticas, Universidad Carlos III de Madrid, 28911 Legan\'es, Madrid, Spain}
\affil{Instituto de Biocomputaci\'on y F\'\i sica de Sistemas Complejos (BIFI), Universidad de Zaragoza, 50018 Zaragoza}

\date{\today}

\maketitle

\begin{abstract}

Network ecology is a rising field of quantitative biology representing ecosystems as complex networks. A suitable example is parasite spreading: several parasites may be transmitted among their hosts through different mechanisms, each one giving rise to a network of interactions. Modelling these networked, ecological interactions at the same time is still an open challenge. We present a novel spatially-embedded multiplex network framework for modelling multi-host infection spreading through multiple routes of transmission. Our model is inspired by \textit{Trypanosoma cruzi}, a parasite transmitted by trophic and vectorial mechanisms. Our ecological network model is represented by a multiplex in which nodes represent species populations interacting through a food web and a parasite contaminative layer at the same time. We modelled Susceptible-Infected dynamics in two different scenarios: a simple theoretical food web and an empirical one. Our simulations in both scenarios show that the infection is more widespread when both the trophic and the contaminative interactions are considered with equal rates. This indicates that trophic and contaminative transmission may have additive effects in real ecosystems. We also find that the ratio of vectors-to-host in the
community (i) crucially influences the infection spread, (ii) regulates a percolating phase
transition in the rate of parasite transmission and (iii) increases the infection rate in hosts. By immunising the same fractions of predator and prey populations, we show that the multiplex topology is fundamental in outlining the role that each host species plays in parasite transmission in a given ecosystem. We also show that the multiplex models provide a richer phenomenology in terms of parasite spreading dynamics compared to more limited mono-layer models. Our work opens new challenges and provides new quantitative tools for modelling multi-channel spreading in networked systems.

\vspace{0.2in}

\end{abstract}

%


\setstretch{1.1}

\vspace{0.2in}

\section{Introduction}
Pathogens and parasites ("parasites" hereafter) are one of the most widespread and
diverse life forms \cite{poulin2014parasite, dobson2008homage}. Several parasites infect multiple host species \cite{rigaud2010parasite} and many of these parasites may infect their host using different routes of transmission \cite{poulin2011evolutionary}.  Multi-host parasites include many zoonoses with complex dynamics that challenge infection control and prevention efforts \cite{dobson2004population}. For instance, several multi-host protozoan parasites of public health concern exhibit more than one mode of transmission: \textit{Toxoplasma gondii} can infect its hosts by fecal-oral transmission, the consumption of an infected prey, and through the placenta \cite{dubey2004toxoplasmosis}; \textit{Cryptosporidium} directly infects its hosts via sexual contact or via fecal-oral transmission \cite{fayer2000epidemiology}; \textit{Trypanosoma cruzi} can be transmitted by insect vectors, the consumption of an infected prey, and also through the placenta \cite{noireau2009trypanosoma, jansen2015multiple}. This complexity of host types and transmission modes challenges the development of models that account for the different sources of variation. The network approach is a promising alternative because it allows accounting for the individual, species-level and spatial sources of heterogeneity \cite{craft2011network,barter2016spots}.

Contact networks can be explicitly used to understand the epidemiological consequences of complex host interaction patterns \cite{keeling2005implications, meyers2005network, bansal2006comparative, ferrari2006network, craft2009distinguishing, dalziel2014contact}. In a contact network, each individual is represented as a node and each contact that potentially results in transmission  between two nodes is represented as a link. Interactions can also be embedded in space \cite{craft2009distinguishing, davis2008abundance, davis2015spatial} where the probability of interaction between nodes may depend on the distance between them. The number of contacts of a node is called the degree of the node and is a fundamental quantity in network theory \cite{dalziel2014contact}. All epidemiological models make assumptions about the underlying network of interactions, often without explicitly stating them. For example, classical mean-field models used in epidemiology assume that all the interactions have the same probability of leading to transmission \cite{anderson1992infectious}. Contact network models, however, mathematically formalise this intuitive concept so that epidemiological calculations can explicitly consider complex patterns of interactions \cite{bansal2007individual}. A different approach consists in considering meta-population dynamics \cite{colizza2008modeling}, instead of individual contacts. 

Recently, the recognition that real-world networks may include different types of interactions among entities prompted the development of methods that take into account the heterogeneity of
interactions as well \cite{boccaletti2014structure, kivela2014multilayer}. Examples include multi-modal transportation networks in metropolitan areas \cite{barthelemy2011spatial,Morris2012transport,lima2015disease}, or proteins that interact with each other according to different regulatory mechanism \cite{Cardillo2013,Cozzo2013}. Ecological systems are also characterised by multiple types of relationships among biological entities, organised and structured on different temporal and spatial scales \cite{kefi2015ecological, kivela2014multilayer}. Different interaction types can be described as "multiplex networks" \cite{wasserman1994social,kefi2015ecological,mucha2010community,battiston2014structural,dedomenico2013mathematical}. Multiplex networks are a particular kind of multi-layer networks where the same nodes appear on all the layers but they can be connected differently on each layer. Each multiplex layer contains edges of a given type. In the context of parasites that can be transmitted over multiple transmission channels, multiplex networks can be used to include distinct mechanisms of parasite transmission  \cite{kefi2015ecological}. This approach encapsulates the heterogeneity in the transmission of real-world diseases and helps us understand how the interplay between different modes of transmission affects infection dynamics in an ecosystem \cite{Buono2014, salehi2014diffusion,lima2015disease}. 

Descriptions of ecological multiplex networks \cite{kefi2015ecological,kefi2015multiplex} and studies of infection spreading over multiplex structures \cite{Buono2014, salehi2014diffusion,gomez2013diffusion} have recently appeared in the literature.  Previous approaches have already described the structural characteristic of food webs that include parasites \cite{lafferty2006parasites} and tried to incorporate parasites in food webs using network framework \cite{lafferty2008parasites}. The effect of multiple hosts on parasite spreading dynamics have also been explored in the context of disease risk \cite{keesing2006effects}, disease emergence in a target host \cite{fenton2005community}, parasite sharing and potential transmission pathways \cite{pilosof2015potential} and also in a multilayer network exploring cross-species transmission (within and between host species) \cite{pilosof2015asymmetric}. However, the consideration of real ecological scenarios in the analysis of parasite spreading through multiple transmission mechanisms is still an open problem. We propose a spatial multiplex-based framework to model multi-host parasite transmission
through multiple transmission mechanisms. In this framework, each transmission mechanism can be represented in a different layer of the multiplex network structure. Our model is inspired by the complex ecology of \textit{Trypanosoma cruzi} (Kinetoplastida: Trypanosomatidae) in its multiple host community. \textit{T. cruzi} is a relevant example of a multi-host parasite and in humans it causes the Chagas disease, a serious infection affecting 6-9 million people \cite{hotez2008neglected}. The main infection route to humans involves the insect vectors (triatomine kissing bugs), but oral transmission is also recurrent \cite{shikanai2012oral}. Vectors get infected when consuming blood meals from an infected host, while host infection occurs through the contact of vector's faeces and the biting wound or mucosa (stercorarian transmission). In sylvatic hosts the stercorarian transmission may occur when the animal scratches the bite and inadvertently rubs the parasite-contaminated matter into the lesion \cite{kribs2006vector}. Infection by the oral route occurs when a mammal host ingests infected triatomine faeces, food contaminated with the parasite or by preying on infected vectors or mammals \cite{jansen2015multiple}.

Preliminary studies \cite{kribs2006vector, kribs2010estimating, pelosse2012role} used mean-field methods to model \textit{T. cruzi} transmission among its hosts and vectors. Their results indicate that in a fully connected network with no explicit spatial structure, vectorial and oral transmission effects are additive in maintaining and furthering the spread of the infection \cite{kribs2006vector}. 
We use a Susceptible-Infected (SI) model to describe parasite transmission dynamics in spatially embedded multiplex networks. The multiplex framework can help us understand how infection spread is related to different ecological interactions and what is the epidemiological importance of vectors and hosts in different ecological scenarios. 
We first investigate the parasite spreading across aggregated parasite-host and trophic interactions. In order to measure the influence of the spatial embedding, we contrast the behaviour of a non-spatial model against one where nodes are embedded in space. 
We then study a reference spatial multiplex network in order to understand the interplay between the multiplex structure and epidemiological dynamics. In the vectorial transmission layer, vectors are contaminated after interacting with infected hosts and transmit the parasite when interacting with non-infected hosts. In the trophic transmission layer hosts acquire the parasite after feeding on infected vector or host. Finally, we use empirical data of a local \textit{T. cruzi} host community, the Serra da Canastra ecosystem \cite{rocha2013trypanosoma}, to model the dynamics of \textit{T. cruzi} multiple transmission routes on its multiple hosts. 

With the multiplex framework we aim to understand the effect of multiplex topology and the relative importance of vectorial and trophic transmission for parasite spreading dynamics. We use multiplex cartography \cite{battiston2014structural} to characterize species structural importance in the network and compare scenarios with different relative frequency of vectors. We then explore the speed of parasite spreading depending on the importance of vectorial and trophic transmission in scenarios with different frequency of vectors. Finally, we explore the effect of species structural importance on parasite spreading by simulating immunisation experiments. 

\section{Methods} 

We model a  set of $N$ populations interacting within an ecosystem via a network framework. Our aim is to model the diffusion of a multi-host parasite within the ecosystem. Nodes represent populations and they have identities, i.e. their species types (predator, prey, and vector). We denote with $S=\{s_k\}_{k=1}^{s}$ the set of all the $s$ species types. Each node in the network is of a given species type $s_k$ with frequency $f_k$, normalised such that $\sum_{k=1}^{s}f_{k} = 1$.

Given that we do not have enough information about the individual-level patterns of interactions, we will consider the food-webs in terms of interacting populations. We consider nodes as populations that follow the same formalism of individual-based dynamics. Our approach is based on the following assumptions: (i) we consider that the parasite transmission is fast and that all the individuals within a population instantaneously gets infected once transmission occurs (in other words, we do not consider meta-population dynamics such as considering parasite spreading within the population and dispersal among populations \cite{colizza2008modeling}); (ii) we consider the parasite spreading happening at a much faster rate than any birth-death dynamics (which we do not consider).

We assume that individuals from populations can disperse across the system and potentially interact with other populations, according to a \textit{dispersal layer}. The dispersal layer is  an undirected graph with adjacency matrix $D$, so that $d_{ij} = d_{ji} = 1$ if population $i$ can interact with $j$ and vice-versa. In the following subsections, we define the topology of the dispersal layer as being either an Erd\"os-R\'enyi random graph or a random geometric graph. The main difference between the two is that the latter includes the notion that only spatially close enough populations can interact with each other (since on random geometric graphs nodes are embedded in space and linked if closer than a certain threshold distance $\rho$).

\begin{figure}
\includegraphics[scale=.36]{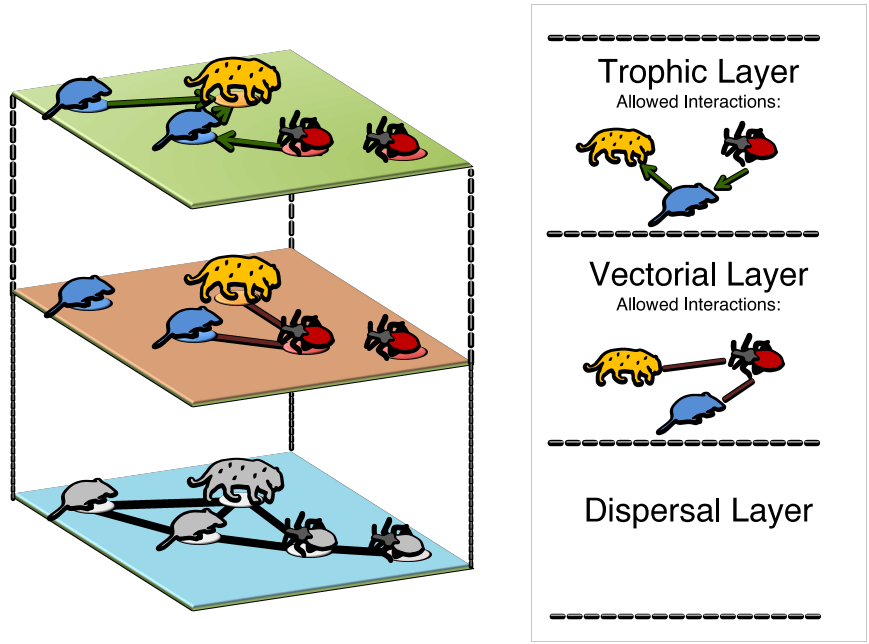}
\caption{Visual representation of our model over the three layers: a trophic layer, a vectorial layer, and their underlying dispersal layer. Nodes are relative to the three-species example and they are drawn according to their species types, e.g. ``predator", ``prey" and ``vector". Trophic and vectorial layers allow only for specific interactions to be present within the system, according to the species types involved in them. For instance, the allowed interactions in the three-species model are reported on the right. The parasite can spread on both such layers. When a node gets infected in one layer it gets infected on all the others as well. While the dispersal layer induces the other two, it is only the trophic and the vectorial layers that constitute our ecological multiplex networks.}  
\label{fig1} 
\end{figure}

In our model, population interaction can potentially give rise to either (i) trophic interactions (a given species feeding on another one) or (ii) contaminative interactions (a given species of host getting in touch with vectors and transmitting the parasite through blood exchanges). Considering only trophic (or contaminative) interactions gives rise to the trophic (or vectorial) layer. Alternatively, considering both interactions together gives rise to an aggregated layer. A visualisation of the dispersal, trophic and contaminative layers is provided in Figure~\ref{fig1}.

Transmission on a given network layer are allowed according to node identities $\{s_k\}$ and are defined according to the corresponding $s\times s$ \textit{interaction matrices}, $T$ for the trophic layer, $V$ for the vectorial layer and $A=T \oplus V$ for the aggregated layer, where $\oplus$ indicates the Boolean OR function. There is no direct interaction between populations of the same species type because there is no cannibalism in the trophic layer and also no parasite transmission among vectors in the vectorial layer. This means the main diagonal of all interaction matrices are all 0s. The sifting of the dispersal layer through either $T$ or $V$ or $A$ produces $s$-partite graphs, i.e. there are no edges between nodes of the same species types. We notice that sifted trophic interactions give rise to a directed network layer while we obtain an undirected vectorial layer from allowed contaminative interactions.

Providing the collection of species types $S$, the topology of the dispersal layer $D$, choosing if considering trophic and vectorial layers as separate or aggregated, and defining the corresponding interaction matrices fully determines the model. 
We explore the following models, enlisted in order of presentation:
\begin{itemize}
\item a random graph as dispersal layer, with 3 species types and aggregated interactions, called Random Aggregated Network (RAN);
\item a random geometric graph as dispersal layer, with 3 species types and aggregated interactions, called Spatial Aggregated Network (SAN);
\item a random geometric graph as dispersal layer, with 3 species types, interactions kept separate  across a 2-layer multiplex structure, called Spatial Multiplex Network (SMN);
\item a random geometric graph model, with 20 species, interactions kept separate across a 2-layer multiplex structure according to ecological empirical interactions. This model is called Spatial Ecological Multiplex Network (SEMN).
\end{itemize}
We considered both smaller ($N=1,000$ nodes) and larger networks ($N=10,000$ nodes) with the same average degree. While the results obtained in both cases were robust to the network size change, the  
networks with $N=10,000$ nodes displayed less finite-size effects. Therefore, in the following we present simulation and analytic results for networked ecosystems made of $N=10,000$ nodes.
The average degree of considered networks has been tuned in order to obtain connected dispersal layers, in which there is at least one path connecting each pair of nodes. This minimises statistical biases due to disconnectedness of a non-negligible fraction of populations. 

\subsection{Random aggregated network model}

In the random aggregated network model (RAN) nodes have $s=3$ possible identities, $S={s_1,s_2,s_3}={\text{predator,prey,vector}}$ with species frequencies $f_1,f_2,f_3$ respectively. Herbivorous mammals are in general more abundant than carnivorous \cite{damuth1981population} and for sake of simplicity we assume prey populations being double as frequent as predator populations, $f_2=2f_1$. Therefore, given that $f_1+f_2+f_3=1$, one obtains that $f_1=(1-f_3)/3$ and $f_2=2(1-f_3)/3$, thus leaving the vector frequency $f_3=f_v$ as a free parameter of the model.
In this model the dispersal layer has the topology of an Erd\"os-R\'enyi with probability $p_{ER}$. Therefore, no space is included in the RAN model. In order to consider fully connected graphs in our simulations and to reduce the effects of degree heterogeneity we fixed a $p_{ER}$ giving rise to networks with average degree $\left\langle k_{ER} \right\rangle = p_{ER}\cdot (N-1) \approx 28.27$. 
The RAN model sifts interactions among predator, prey and vector populations from the dispersal layer according to the interaction matrix $A$ defined as:

\begin{equation}
A=T\oplus V=\left(\begin{array}{ccc}
0 & 0 & 0\\
1 & 0 & 0\\
0 & 1 & 0
\end{array}\right)\oplus\left(\begin{array}{ccc}
0 & 0 & 1\\
0 & 0 & 1\\
1 & 1 & 0
\end{array}\right)=\left(\begin{array}{ccc}
0 & 0 & 1\\
1 & 0 & 1\\
1 & 1 & 0
\end{array}\right).
\label{eq:matrix}
\end{equation}

For instance, $t_{21} = 1$ means that $s_2=\text{prey}$ populations are eaten by $s_1 =$ predator populations. Notice that allowed interaction in $T$ are directed (from the eater to the eaten, as usual in food-webs \cite{Bueno2003,Ramos2007,kefi2015ecological}) while they are undirected in $V$, since they represent ecological exchanges of infected fluids between the host and the vector species \cite{rocha2013trypanosoma}).
The above sifting creates the aggregated single layer of the model, where trophic and contaminative interactions are combined and where parasite diffusion occurs. 

\subsection{Spatial aggregated network model}

In the spatial aggregated network model (SAN) the dispersal layer is a random geometric graph (RGG). Therefore, populations are embedded in a space. Nodes are scattered uniformly at random within the 2D space $\Omega=[0,1]^2$ with periodic boundary conditions, i.e. a toroidal space. As known from previous works \cite{sattenspiel2009geographic}, the average degree of an RGG is $\left\langle k_{RGG} \right\rangle = \pi N \rho^2$. For the sake of comparisons with the RAN model, we chose $\rho = 0.03$, thus having $\left\langle k_{RGG} \right\rangle = \left\langle k_{ER} \right\rangle = 28.27$.
The interaction matrix $A$ sifting the only aggregated network layer is the same as in the RAN model. Also species types are distributed as in the RAN model.

\subsection{Spatial multiplex network model}

In the spatial multiplex network model (SMN) the dispersal layer is a random geometric graph (RGG) with nodes spatially embedded and species types distributed  as in the SAN model. 
However, we keep trophic and contaminative interactions as distinct on two separate layers. These structured interactions give rise to a multiplex network \cite{kivela2014multilayer, kefi2015ecological,boccaletti2014structure}, where populations are replicated across both layers and no explicit inter-layer edges are considered \cite{dedomenico2013mathematical}.
The interaction matrices sifting the trophic and the vectorial layer are respectively $T$ and $V$, as defined above in Equation~\ref{eq:matrix}. 
A multiplex network visualisation of the SMN model is provided in Figure~\ref{fig1}.

\subsection{Spatial ecological multiplex network model}

In our last model, the spatial ecological multiplex network (SEMN), the dispersal layer is a random geometric graph (RGG), as in the SAN model. Also, trophic and contaminative interactions are kept separate analogously to the SMN model. In SEMN we used empirical ecological data within the model~\cite{rocha2013trypanosoma}.
Specifically, we use data from an epidemiological study of \textit{T. cruzi} infection in wild hosts in Southeast Brazil \cite{rocha2013trypanosoma} to estimate the trophic and vectorial interaction matrices $T_{eco}$ and $V_{eco}$ (see Supplementary Information), considering a total of 20 species. 
For the trophic interaction matrix $T_{eco}$, we build a qualitative potential food-web based on the animals diets \cite{Bueno2003,Ramos2007,Cavalcanti2010,Amboni2007,CarvalhoNeto2012,Reis2011}. As there was no species-level classification of the biological vectors present in the area, we considered the vectors as one single species type. We use species prevalence to estimate contaminative interactions in $V_{eco}$ \cite{rocha2013trypanosoma}. We assume that positive parasitological diagnostics for \textit{T. cruzi} could be used as a proxy for vectorial transmission, since only individuals with positive parasitaemia (i.e. with high parasite loads in their blood) are able to transmit the parasite \cite{jansen2015multiple}. The vectorial layer was constructed based on the assumption that species with positive prevalence in hemocultive transmit the parasite to vectors and that species with positive prevalence in serology can be infected from vectors.
The SEMN model has a total of 20 species types: $a=7$ predators, $b=12$ prey and 1 vector species. As in the previous models, we assumed that prey populations have double the frequency of predator populations (see RAN model). We considered all the predator and prey species populations having identical frequencies $f_{pred}$ and $f_{prey}$ respectively, such that:
\begin{equation}
 a f_{pred} + b f_{prey} + f_v = 1 \rightarrow f_{prey} = 2\frac{1-f_v}{a+2b} = 2f_{pred}.
\end{equation}
Therefore, by tuning $f_v$ we change also the frequency of predator and prey populations.
The SEMN model is the most realistic one of this study since it takes into account spatial embedding, multiplex structure and empirical ecological data.

\subsection{Parasite transmission dynamics}

To simulate the parasite transmission dynamics a node, i.e. a population of a given species type can be either susceptible or infected. We start the simulation by infecting a fraction $\phi_0=0.28\%$ of all populations. In the RAN model we infect one node at random and let the infection spread along a random walk on the dispersal layer. We start measuring the infection dynamics after $N\phi_0$ nodes are infected. Similarly, in the other three spatial models, we infect 
all the nodes in a random circle of radius $r_0=0.03$, that is, $\pi Nr_0^2 \approx 28.2$ populations become infected at the beginning, on average (a sensitivity analysis proves that the results presented in the following sections are robust up to 5\% of initially infected populations). 
Subsequently, the parasite spreading evolves in SMN and SEMN models as follows:
\begin{enumerate}
\item A random node $i$ is chosen together with one of its neighbours $j$ on the dispersal layer.
\item The vectorial layer is chosen to be considered for the parasite transmission with probability $p_v$, which is a measure of the vectorial layer importance. Step 3 is then performed when the vectorial layer is chosen. Otherwise, step 4 takes place.
\item If node $i$ is infected and the edge $(i,j)$ exists in the vectorial layer, 
 node $j$ becomes infected as well (vectorial layer parasite transmission).
\item If node $i$ is infected and the edge $(i,j)$ exists in the trophic layer, 
 node $j$ becomes infected as well (trophic layer parasite transmission).
\item Steps 1-4 are repeated $N=10^4$ times per each time step, i.e. an average of 1 update per node per time step, until $T_{max}$ time steps are reached.
\end{enumerate}

For RAN and SAN models parasite transmission occurs only on the aggregate layer without considering steps 2, 3 and 4. This is equivalent in treating contaminative and trophic interactions in an aggregate, unweighted way. Each population can be randomly chosen at each time step and at the end of the transmission process every node is chosen once, on average. 
This parasite transmission model is equivalent to an SI model with contact rate $\beta=1$, where $\beta$ is the probability for an individual to become infected when exposed to the disease \cite{sattenspiel2009geographic}. For the sake of simplicity, we assume $\beta=1$ in both the trophic and vectorial layers and across all the species. Our assumption leads to the disease firstly spreading across the geodesic paths of the multiplex topology \cite{sattenspiel2009geographic,lucas2015geodesics} so that our infection process depends solely on the multiplex network structure. Notice that a more complicated model with two different $\beta$, one for each layer, would still be expected to reproduce a similar phenomenology to the one reported in the following (with one $\beta$ only). This is because our infection dynamics is an SI model and because even the simpler model with one $\beta$ only still potentially weights differently each layer through $p_v$.

\subsection{Model parameter values}

Let us summarise the main parameters of our models and relative values. In this study we consider networks of $N=10,000$ populations (nodes) and average degree $\left\langle k \right\rangle = 28.27$ for the dispersal layer ($p_{ER}=\left\langle k \right\rangle/(N-1)$ for random graphs, $\rho=0.03$ for RGGs). We chose these parameter values in order to consider fully connected multiplex networks. Let us underline that we consider a multiplex connected component as the set of all nodes that can be reached from each other by considering all edge types of a node~\cite{dedomenico2014navigability}. Given that we have directed edges in the trophic layer, we have to consider the notion of \textit{strongly connected component} on the multiplex topology, i.e. a set of nodes that can be reached from each other considering oriented paths along directed edges of any colour.

The maximum number of time steps $T_{max}=10^4$ has been numerically tuned in order to let the system reach equilibrium. Each time step considers $N=10^4$ updates for the parasite spreading dynamics, i.e. an average of 1 update per node per time step.
The frequency of vector populations $f_v$ is a free parameter of the model, together with the vectorial layer importance $p_v$, i.e. the probability for the parasite to spread along the vectorial layer, in the SMN and SEMN models.

\subsection{Immunisation}

In order to investigate the role played by predators and prey populations in spreading the parasite we focus on multiplex models (SMN and SEMN models). Using immunisation simulations we study the dynamics of parasite spreading when the same number of either predator or prey populations have been immunised. An immune node is not susceptible to the parasite. The number of immune nodes is determined per species by specifying the probability of immunisation $\pi_k$ for each species $k\in S$. To perform the immunisation, populations of species $s_k$ are randomly chosen with probability $\pi_k$ and are set to be immune.   

We consider two immunisation scenarios to investigate the relative role that predator or prey populations have in spreading the parasite. In the first scenario only prey populations are immunised while in the second scenario only predator populations are immunised. For simplicity, the $\pi_k$ values for all prey and predator populations are set uniformly, however they are chosen in order to immunise the same total number of predators and the same total number of prey. From an ecological point of view, the immunisation simulations answer the following question: given the fictional possibility of vaccinating a limited number $\phi \ll N$ of populations against the parasite, is it more efficient to immunise predator populations or prey ones in order to hinder the parasite spreading?

\subsection{Multiplex cartography}

A multiplex cartography visually represents the role played by a given node across different layers according to its topological features \cite{guimera2005cartography,battiston2014structural}. In this way, multiplex cartography becomes a rather simple yet powerful network metric providing information on the topological patterns of nodes across the multiplex structure. We chose it compared to other multiplex measures because of its simplicity, its powerfulness and its appealing analogy with maps. We build on previous literature~\cite{dedomenico2013mathematical,battiston2014structural} by considering a cartography based on the following two measures: the multidegree or overlapping degree $o_i$ and the participation coefficient $P_{i}$ of node $i$. As in~\cite{battiston2014structural,dedomenico2013mathematical}, the multidegree $o_{i}$ is defined as the sum of all the degrees of node $i$ across the $M$ multiplex layers:

\begin{equation}
o_{i}=\sum_{\alpha}k_{i}^{(\alpha)}.
\end{equation}
where $k_{i}^{(\alpha)}$ is the degree of node $i$ in the layer $\alpha \in \{1,...,M\}$. 
The overlapping degree $o_i$ represents a proxy of the overall local centrality that a node has within the multiplex network. Differently from~\cite{battiston2014structural}, we consider $o_i$ rather than its standardised counterpart $z_i=\frac{(o_i - \langle o_i \rangle)}{\sigma(o_i)}$ because our multiplex networks do not display Gaussian-like multidegree distributions. We consider hubs in our multiplex networks as those nodes being in the 95th percentile of the multidegree distribution. 

The distribution of the connections over the different layers can be expressed via the participation coefficient $P_{i}$ of node $i$:

\begin{equation}
P_{i}=\frac{M}{M-1}\left[1-\sum_{\alpha=1}^{M}\left(\frac{k_{i}^{(\alpha)}}{o_{i}}\right)^{2}\right].
\end{equation}

$P_{i}$ ranges between 0 (for nodes that concentrate all their connections in one level only) and 1 (for nodes that distribute connections over all the $M$ layers uniformly). 
In the following, we visualise our multiplex network cartography by clustering together individual points (each one referring to a given node) into 2D bins, thus obtaining a 2D histogram resembling a heat-map. The binned quantities are the overlapping degree on the y-axis and the participation coefficient on the x-axis. 

\subsection{Infection measures}

On a macroscopic scale, we investigate parasite spreading by computing the \textit{global infection time}, defined as the time step at which the largest (in node size) weakly connected component of the multiplex network is infected. Alternatively, the infection time indicates the time step $t_{inf}$ at which the disease infects most of the nodes within the network. If $R(t) = N_{inf}(t)/N $ if the ratio of infected populations/nodes at time $t$, then $\text{Max}_{t}(R(t))=R(t_{inf})$. 

Infection times represent a global, macroscopic statistics of the parasite spreading. To analyse the evolution of transmission  in more detail we use the \textit{parasite ratio increase} $\Delta R(t) := R(t+1)-R(t)$, i.e. the increase of the ratio of infected populations in one time step. The $\Delta R(t)$ is a measure for the rate at which the parasite is spreading within the multiplex network.

In order to capture the spatial features of our SMN and SEMN models we measure also $\langle \lambda \rangle$ defined as the average distance of the infected nodes from the centre of the embedding square $\Omega:=[0,1]^2$ (where the infection originates). Given our assumption of uniform spreading of species populations within $\Omega$, it is relatively straightforward to compute an upper bound $\langle \lambda \rangle^* $ for $\langle \lambda \rangle$ as:

\begin{equation}
\langle \lambda \rangle^*  = \iint_{0}^{1} \sqrt{(x - \frac{1}{2})^{2}+(y - \frac{1}{2})^2} dxdy \approx 0.3826.
\end{equation}

$\langle \lambda \rangle^* $ represents the maximum average distance of infected populations from the centre of the embedding space (also the origin of the infection).

\section{Results} 

Our results focus on: (i) highlighting the role of spatial correlations on the parasite spreading dynamics, (ii) assessing the differences between aggregated and multiplex models, (iii) highlighting the topological features of our models through cartography \cite{battiston2014structural} while relating them to parasite spreading at different values for the vector frequency $f_v$ and importance of vectorial transmission $p_v$, and (iv) quantifying how different species promote or not parasite spreading by means of immunisation simulations. We first report the results concerning the aggregate models (RAN and SAN), then the three-species reference one (SMN) and the spatial ecological multiplex network (SEMN) as last. 
In particular, we show that: (i) the presence of spatial correlations slows down the parasite spreading in the SAN model compared to the RAN one, (ii) the multiplex structure deeply influences the parasite spreading dynamics in both SMN and SEMN models, (iii) the vector frequency determines a percolation threshold in the parasite spreading rate over the whole networked ecosystem in both SMN and SEMN models, (iv) a higher biodiversity in the SEMN model significantly modifies the infection times in similarly sized ecosystems from the SMN model and (v) prey and predator populations play different roles in promoting the parasite spreading in the empirical SEMN scenario.

\subsection{Aggregate network models: the role of space}

\begin{figure}
\centering
\includegraphics[width=.55\paperwidth]{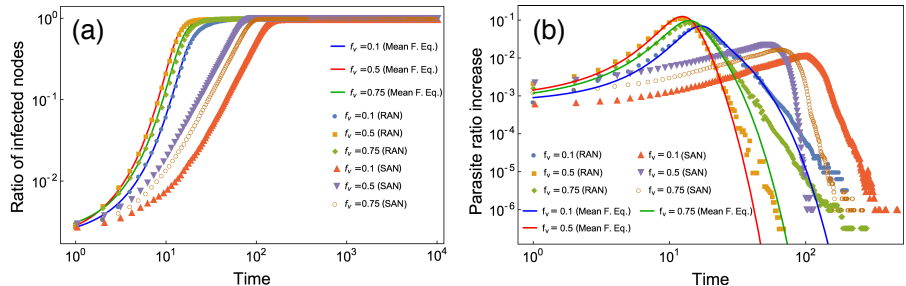}
%
\caption{(a): Ratio of infected nodes over time for the random aggregate network (RAN) and the spatial aggregate network (SAN) models, at different frequencies $f_v$ of vector populations in the system. (b): Parasite ratio increase of infected nodes over time for the random aggregate network (RAN) and the spatial aggregate network (SAN) models, at different frequencies $f_v$ of vector populations in the system.
}  
\label{fig2} 
\end{figure}

Comparing the results of the aggregate models RAN and SAN provides quantitative information about the role played by space.
In Figure \ref{fig2} (a) we compare the ratio of infected nodes over time for the RAN and SAN models by means of simulations and analytical results. Assuming a mean-field approximation, where every population can be potentially infected by any other one in the system, it is possible to write down the following equations for the infection dynamics:

\begin{equation}
\centering
\dot{n}_{1}=f_{1}N\left(\frac{f_{1}N-n_{1}}{N}\right) \left(\frac{n_{2}}{N}+\frac{n_{3}}{N}\right)
\label{equa:pred}
\end{equation}

\begin{equation}
\centering
\dot{n}_{2}=f_{2}N\left(\frac{f_{2}N-n_{2}}{N}\right) \frac{n_{3}}{N}
\label{equa:prey}
\end{equation}

\begin{equation}
\centering
\dot{n}_{3}=f_{3}N\left(\frac{f_{3}N-n_{3}}{N}\right) \left(\frac{n_{1}}{N}+\frac{n_{2}}{N}\right)
\label{equa:vect}
\end{equation}

where $n_k=n_k(t)$ is the number of infected nodes of species type $k\in {1,2,3}$ at time $t$. Each equation considers how a given susceptible species population can be potentially infected in the model through its edges with other species population types. For instance, let us consider the infection dynamics of predator populations ($k=1$). At time $t$, the probability of finding a susceptible predator population in the system is $(Nf_1 - n_1)/N$. However, in all models which consider 3 species, a susceptible predator population can receive the parasite infection either from feeding on infected prey populations (the probability of sampling one is equal to $n_2/N$) or from being contaminated by an infected vector population (the probability of sampling one is equal to $n_3/N$). Analogous reasoning leads to the Equations~\ref{equa:prey} and~\ref{equa:vect}. Notice that having directed edges leads to prey getting infected only through infected vectors in Equation~\ref{equa:prey}.

Even though the mean field approximation does not consider the networked structure of the underlying dispersal layer, Figure \ref{fig2} (a) shows that analytical results from the mean field equations reasonably approximate simulation results on ER random graph topologies (in RAN) at different vector frequencies $f_v$. Theory and simulations agree in indicating that the infection spreading dynamics reaches its maximum value around 20 time steps in the RAN model. Increasing the vector frequency does not always lead to the infection dynamics reaching its maximum value in less time steps. In fact, when we have $f_v=0.1$ the ratio of infected nodes reaches its maximum value later than in the $f_v=0.5$ case, i.e. the global infection time decreases. However, further increasing vector frequency  from $f_v=0.5$ to $f_v=0.75$ leads to an increase rather than to a reduction in the global infection time.

For completeness, we also show in Figure~\ref{fig2} (b) the relative parasite ratio increases indicating the rate of parasite diffusion over time. We notice that the RAN model always displays a peak over time in the parasite ratio increases. This means that the parasite diffusion initially accelerates and it later slows down since susceptible populations become rarer in the system. Simulations and analytical results for the RAN model also agree in the appearing ordering of these peaks. Here, reaching earlier the maximum ratio of infected nodes means reaching earlier the peak in the parasite ratio increase.
This is because we assume that populations of the same species type do not interact with each other (i.e. our networks are $k-$partite graphs). Since infection must always pass through a vector-host-vector path in order to infect other vectors, adding too many vector populations is detrimental for the global infection time.

In the SAN model, when the dispersal layer changes from an ER random graph to an RGG, the infection reaches its maximum spread at a much later stage (around 100 time steps). We observe that inserting spatial correlations makes the mean field approximation unreliable in describing the simulation results. This is due to the spatial embedding giving rise to non-negligible correlations among nodes.

Parasite ratio increases reveal that the RAN model displays also a faster infection spreading dynamics when compared to its spatial counterpart, the SAN model. Interestingly, both the aggregated models display a peak in the evolution of the parasite ratio increases. Overall, the addition of space increases the global infection time and it reduces the parasite spreading rate.

\subsection{Spatial multiplex network model: the role of trophic and contaminative interactions}

The 3-species reference model (SMN) consists of the simplest epidemiological scenario for the multiplex transmission. It is based on the simplest trophic chain in which vectors are consumed by prey populations and prey are consumed by predator populations. In the vectorial layer the vectors contaminate both prey and predator populations, see also Figure~\ref{fig1}. 

\begin{figure}[h!]
\centering
\includegraphics[width=.55\paperwidth]{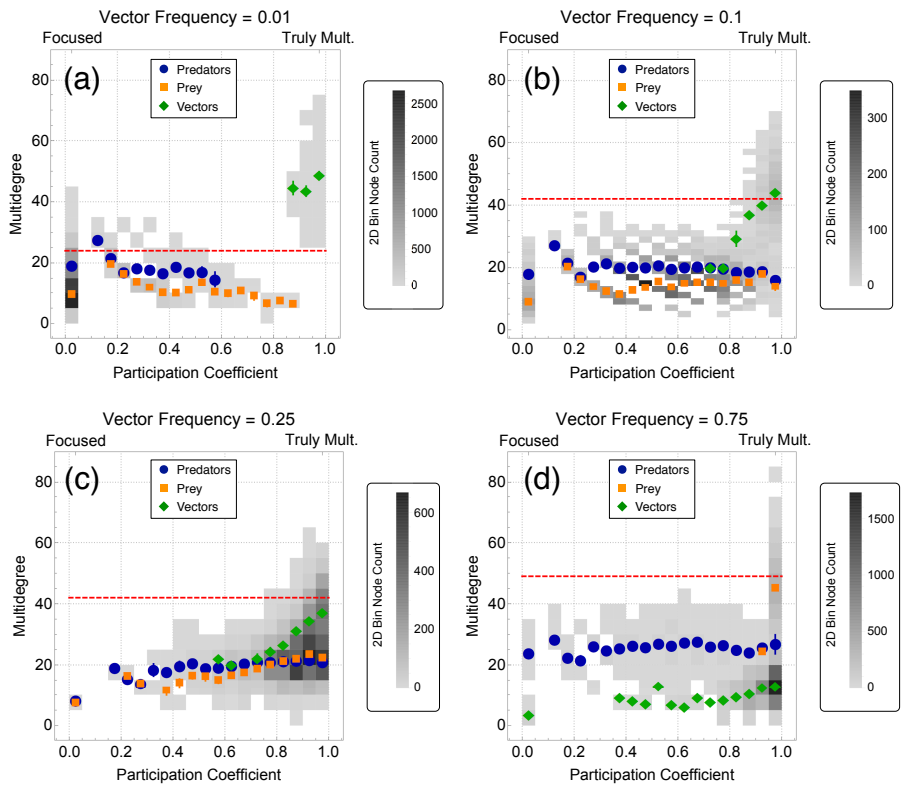}
%
\caption{Cartographies as 2D histograms for the SMN model for vector frequency $f_v=0.01$ (a), $f_v=0.1$ (b), $f_v=0.25$ (c), and $f_v=0.75$ (d). The 10000 multiplex nodes are binned in 2D bins, according to their coordinates in the cartography. Bins are colour-coded according to the number of points falling within them: more coloured tiles have the most nodes in them. Coloured dots identify individual species: predators (blue), prey (orange) and vectors (green). Nodes falling above the horizontal red line have degrees above the 95th percentile in the multidegree distribution and they are therefore considered hubs. Error bars represent standard error of the mean.}  
\label{fig3} 
\end{figure}

In Figures~\ref{fig3} (a)-(d), the multiplex cartographies highlight the degree centrality and participation coefficient of each species type at different vector frequencies $f_v$. Individual nodes are binned according to colour-coded two dimensional tiles so that the resulting plot resembles a heatmap.

When vector populations are rare in the system ($f_v = 0.01$, Figure~\ref{fig3}~(a)), predators' participation coefficient is low. This means that predators interactions are concentrated mostly in the trophic layer and predator populations interact mostly with prey populations. Prey populations show a broader range of participation and this indicates that prey interact with predators and vectors on both layers. Vector populations have the highest participation coefficient and are hubs in the multiplex, since their links are uniformly distributed between both layers. 

When $f_v$ goes from $0.1$, Figure~\ref{fig3} (b), to $0.25$, Figure~\ref{fig3}(c), vector populations show a broader range of participation coefficients indicating that their connections are distributed on both layers. Similar behaviour is reported when $f_v = 0.5$ (plot not presented). At vector frequency $f_v = 0.75$, vector populations are the most frequent in the system and each species type occupies a different region in the cartography (Figure~\ref{fig3}~(d)). Thus, we have:
(i) prey populations linked to vectors on both trophic and vectorial layers becoming almost truly multiplex hubs (participation coefficient value close to one and high multidegree), (ii) predator populations with a broad range of participation coefficients, (iii) vector populations with a broader range of participations coefficients but loosely connected to other populations because vectors do not interact with each other.

\begin{figure}
\centering
\includegraphics[width=0.55\paperwidth]{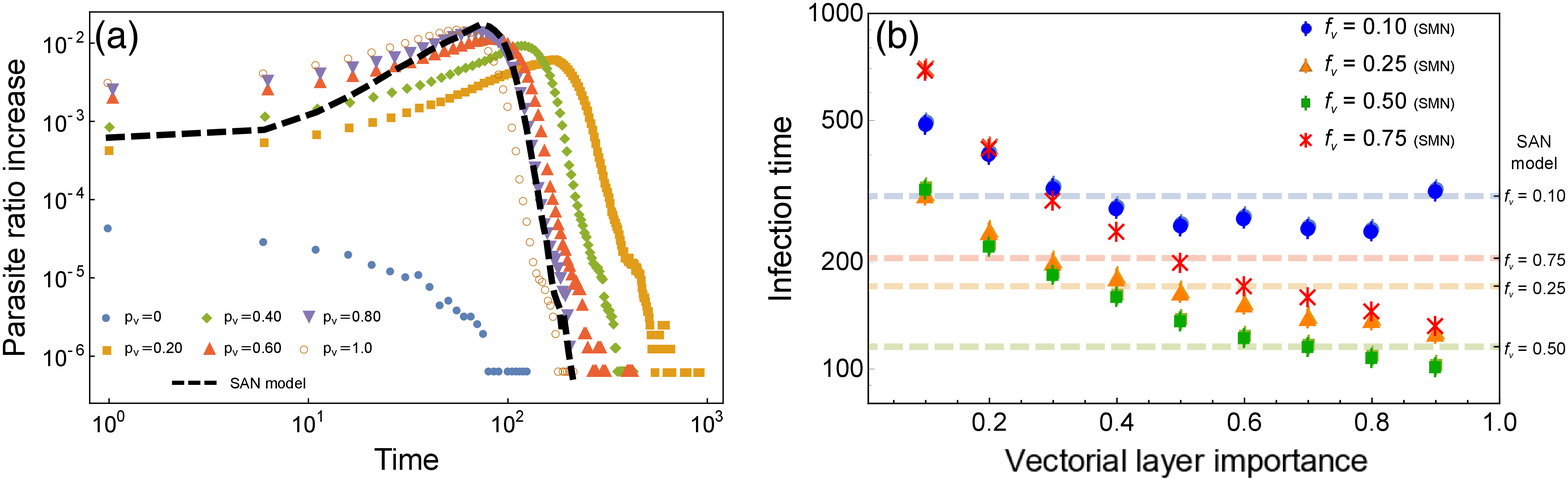}
%
\caption{(a): global infection rate over time for $f_v=0.75$ expressing the diffusion speed of the disease over time for SMN model. A qualitatively similar behaviour was observed also for other vector frequencies. (b): global infection time versus vectorial layer importance $p_v$ for different vector frequencies in the SMN model. Dotted lines represent infection time in the SAN model for different vector frequencies. Results in both plots are averages of 100 repetitions.
}  
\label{fig4} 
\end{figure}

The multiplex structure in the SMN model allows for the infection to spread either on the vectorial layer (with probability $p_v$) or on the trophic layer (with probability $1-p_v$) at each time step (see section 2.5). This interplay leads to the global infection time potentially being a function of the vectorial layer importance $p_v$. As reported in Figure \ref{fig4} (b), when vector frequency is $f_v = 0.1$, the global infection time has its minimum for $0.4<p_v<0.8$. Hence, when the parasite spreads across both trophic and contaminative edges with roughly the same probability, its spreading on the whole multiplex networked ecosystem requires less time. Since the trophic layer in the SMN model is not fully connected and thus the infection cannot reach the entire network, we do not show infection times for $p_v=0$. On the other hand, we do not consider the $p_v=1$ case in order to always consider the food-web while focusing on the multiplex structure.

Increasing the frequency of vector populations does not accelerate parasite spreading in the multiplex network and the faster spreading occurs when $f_v = 0.5$ (\ref{fig4}). The infection time decreases monotonically with the increase of vectorial layer importance $p_v$ when $f_v = 0.25, 0.5$ or $0.75$, but this pattern was not observed when $f_v = 0.1$. This is related to the topology of the allowed interactions in the SMN vectorial layer. In SMN the vectorial layer is undirected and vector populations are connected to both predator and prey populations. The trophic layer has directed interactions and parasite transmission requires at least two steps to spread from vector to predator populations. These topological features of the SMN model enables a faster parasite transmission on the vectorial layer rather than on the trophic layer. However, the frequency of different species types also influences parasite transmission in the model. Increasing the vector frequency from $f_v = 0.1$ to $0.25$ or even up to $0.5$ leads to an overall decrease of the infection times, depending on $p_v$. This trend changes when vectors are the most frequent species type in the system ($f_v = 0.75$). When the majority of nodes are vector populations the speed of parasite spreading increases in relation to  $f_v=0.5$ because vectors are not directly connected in neither of the layers. Therefore, a smaller number of predator and prey populations constraints parasite transmission to vectors. 
In Figure \ref{fig4} (b) we also show the infection time for the SAN model represented as dotted lines for the different vector frequencies. We remember that in the SAN model there is only one aggregated layer where the infection spreads, thus all edges have the same importance. Comparing the infection time of the SAN and SMN models highlights the effect of  multiplex structure for parasite spreading dynamics. Independently on the vector frequency, tuning the parasite spreading across trophic and contaminative interactions changes the infection time with respect to the aggregate case.

The speed of parasite spreading across the multiplex structure also reveals interesting patterns. As reported in Figure \ref{fig4} (a) for $f_v = 0.75$, when $p_v>0$ the parasite transmission initially accelerates within the system ($t<100$). This behaviour is somehow similar to the one already observed in the SAN model (see Figure \ref{fig2} and the black line in Figure \ref{fig4} (a)). On the other hand, when the infection spreads only on the trophic layer ($p_v=0$) a qualitatively different behaviour is observed, with no acceleration phase. This is because of the trophic layer topology (see $T$ in the Methods section): the parasite can spread only from vectors to prey and from prey to predator populations. As the infection spreads, it becomes increasingly difficult to infect more populations over time. Vector populations which are susceptible at the beginning will never be infected. The aggregated model (SAN) does not capture this trend since it includes trophic and contaminative interactions mixed together. We observed a consistent behaviour for other vector frequencies $f_v\neq 0.75$. The only difference was in the order of the peaks of parasite spreading rate: the higher $p_v$ the sooner the peak is reached when $f_v>0.2$. We conjecture that this is because, in environments with many vector populations, the parasite spreads at a faster rate with respect to the trophic layer, so that increasing $p_V$ accelerates the parasite spreading. 
   
\begin{figure}
\centering
\includegraphics[width=0.55\paperwidth]{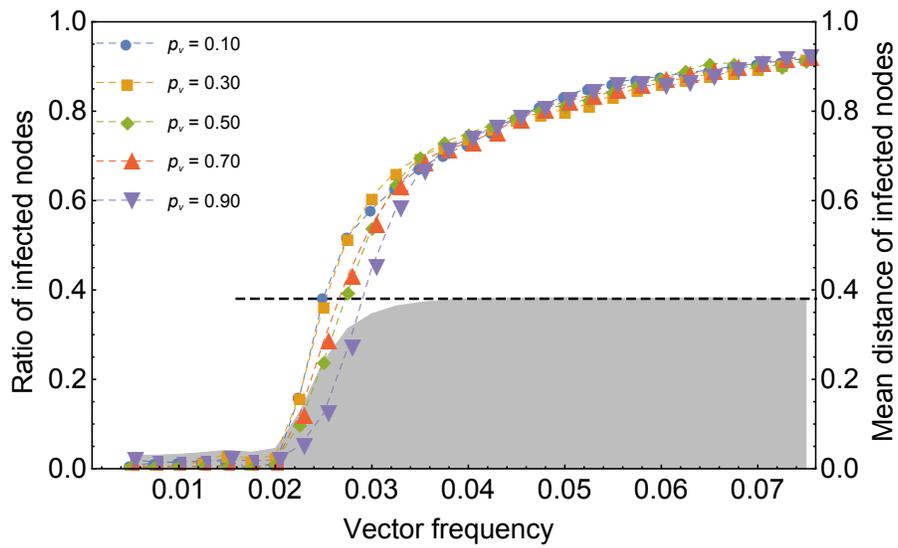}
%
\caption{Ratio of infected populations after $10^4$ steps, sampled at different values of $p_v$, against vector frequency $f_v$ in the SMN model. When vectors are rare in the system, the system displays a phase transition in the rate of infection. The critical threshold is localised around $f_v \approx 0.02$, for all the values of $p_v$. The grey shape represents the mean distance of infected population from the origin of the parasite spreading and it is averaged over different $p_v$ values. When $f_v > 0.02$ the infection radius saturates faster than the global percentage of infected populations. All curves are averages of 100 repetitions.}  
\label{fig5} 
\end{figure}

We also investigated the infection dynamics for very small values of vector frequencies (Figure \ref{fig5}). Simulations indicate that the SMN model displays a critical threshold in the emergence of pandemics around $f_v \approx 0.02$. Very small variations in the abundance of vector populations within the simulated ecosystem leads to dramatic changes in the ratio of infected populations after a suitably long relaxation time of $10,000$ time steps (Figure~\ref{fig5}). By simulating larger ecosystems for $N=25,000, 50,000, 100,000$ and $150,000$, we extrapolated the scaling behavior of the critical threshold of vector frequency $f_v$. Simulation results suggest that the threshold does indeed not vanish in the thermodynamic limit (i.e. $N \rightarrow \infty$) but it is rather close to $f_v \approx 0.02$ and lower bounded by the value $f_v = 0.019$. We conjecture that this critical transition is due to vector populations being fundamental in infecting prey populations. Considering the sifting matrices $T$ and $V$, prey populations can be infected only by interacting with infected vector populations. When vectors are very rare in the system, prey populations (that are quite frequent in the system) get infected at a much slower rate. This bottle-neck translates into a phase transition in the infection rate. Our simulations show that the vectorial layer importance $p_v$ slightly shifts the critical threshold of the phase transition, which occurs across all the different values of $p_v$ (for $p_v=0$ or $p_v=1$ plots not reported for clarity). This phase transition marks the beginning of a distinct ``phase" of the model ($f_v > 0.02$), for which the parasite percolates throughout the whole system at a faster rate, even when vector frequencies are low. Notice that when $0.02<f_v<0.1$, vector populations are multiplex hubs (see (a) and (b) in Figure \ref{fig3}), therefore they promote the parasite spreading on both the SMN layers. 

As indicated by the grey area in Figure \ref{fig5}, the mean distance of infected nodes $\langle \lambda \rangle$ after 10,000 time steps also undergoes a phase transition around $f_v = 0.02$. However, $\langle \lambda \rangle$ converges to its upper bound $\langle \lambda \rangle^*$ at a faster rate compared to the ratio of infected population. Let us consider the case $f_v = 0.04$. The relative ratio of infected nodes is $\approx 70\%$ (see dotted lines in Figure \ref{fig5}), variations in the vectorial layer importance provide no evident fluctuations. However, always at $f_v = 0.04$, the mean distance of infected populations from the centre of infection is not 70$\%$ of the maximum value, but rather $\langle \lambda \rangle (f_v = 0.04) \approx \langle \lambda \rangle^* \approx 0.384$ (see the grey shape and the dashed black line in Figure \ref{fig5}). Therefore, in the same time steps, the infection spreads only across 70\% of populations but it covers almost all the distances from the infection origin, in the embedding space. We interpret this as the parasite spreading at a faster rate uniformly over the whole embedding space rather than uniformly across all the considered populations. These different spatial and number diffusion rates are relative to our selected SI dynamics. When the infection probability $\beta = 1$ (as in our case) and only one neighbour node becomes infected at a time, the infection spreads firstly through geodesics in the network \cite{sattenspiel2009geographic,lucas2015geodesics}. Having the parasite spreading on geodesics through our spatial multiplex network is compatible with our finding from Figure \ref{fig5}: the mean distance of infected nodes from the infection centre saturates faster than the ratio of infected nodes.

\subsubsection{Immunisation scenarios in the SMN model}

\begin{figure}
\centering
\includegraphics[width=.55\paperwidth]{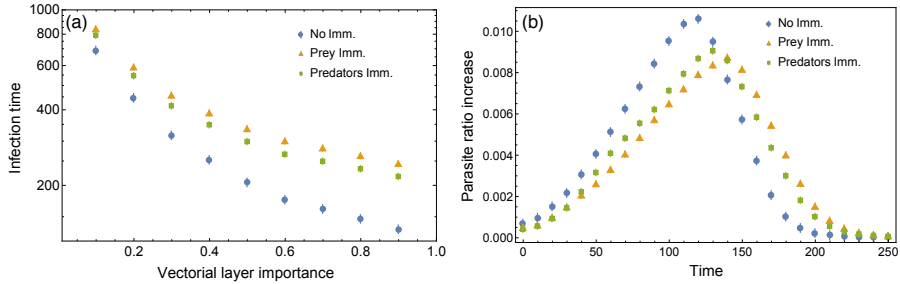}
%
\caption{(a): global infection time versus vectorial layer importance $p_v$ for different immunisation experiments with $f_v=0.75$ in the SMN model. No immunisation means that no immunised populations are present in the system while two other dot types represent scenarios in which only prey or predators are immunised, respectively. For immunisation scenarios the same number of populations has been immunised. (b): parasite ratio increase of infected nodes over time for the SMN model for different immunisation scenarios with $f_v=0.75$. Error bars are computed over 10 independent experiments. Immunising prey is the best choice in terms of both reducing the global infection time and slowing the infection spread over time.}  
\label{fig6} 
\end{figure}

In order to relate the topological features of each species population in the multiplex to their roles in spreading the parasite across the networked ecosystem, we analyse immunisation scenarios. In the immunisation scenarios a fraction of populations of a given species type (e.g. predators) is immunised against the parasite (see Section 2.7). 
As reported in the previous section, we found different species having different degree and participation patterns within the SMN model (see the cartographies in Figure~\ref{fig3}) at high vector frequencies ($f_V = 0.75$). In fact, when $f_v=0.75$ prey, predator and vector populations occupy different regions in the multiplex cartography. 
In Figure~\ref{fig6} we report the global infection times when the same total number $\phi=417$ of predator or prey populations is immunised. The chosen $\phi$ corresponds to immunising half the predator populations in the system. Our results show that immunising prey over predators leads to a greater increase in the system infection times for all values of vectorial layer importance $p_v$. The better performance of immunising prey over predators is also reflected in the increase of parasite ratio $\Delta R(t) $ (Figure \ref{fig6}): immunising prey not only delays a pandemic but it also significantly slows down the parasite spreading in the initial accelerating phase (i.e., it lowers the $\Delta R(t) $ when $t<140$). Even though slowing down the parasite transmission and reaching a pandemic at a later stage might sound equivalent, the parasite ratio increase reveals that in the predator immunisation scenario there is a higher diffusion speed in the decelerating infection phase,  $t>140$ (Figure \ref{fig6}). Because of this behaviour, we report on both  patterns. 

This difference could be attributed to the different topology of prey and predator populations in the trophic layer, i.e., the parasite spreads from vector to prey and then from prey to predator populations, so that prey have a higher betweenness in the sifted trophic interactions. Further numerical experiments indicate that this is not the case. Immunisation experiments performed with the same $\phi$ but with vector frequency $f_v=0.25$ show that immunising either predators over prey gives statistically equivalent results in terms of both the parasite spreading times and the propagation rates. Therefore, at $f_v=0.25$ immunising one species type over the other does not change parasite spreading. However, both the $f_v=0.25$ and the $f_v=0.75$ instances are relative to the same interaction matrices $T$ and $V$ and to the same number of immunised prey $\phi$. Therefore, the relative difference in immunisation performances has to be attributed to the role played by each species within the global network topology. Immunising prey is different from immunising predator populations only when they have different topological patterns within the multiplex network, i.e. they occupy different areas of the multiplex cartography. This evidence points to the meaningfulness of the concept of network cartography for the parasite spreading dynamics: at  $f_v=0.75$ prey populations become truly multiplex hub nodes and assume an important role for parasite spreading, as demonstrated by our immunisation experiments.

\subsection{Spatial ecological multiplex network model: the role of biodiversity}

\begin{figure}
\centering
\includegraphics[width=.55\paperwidth]{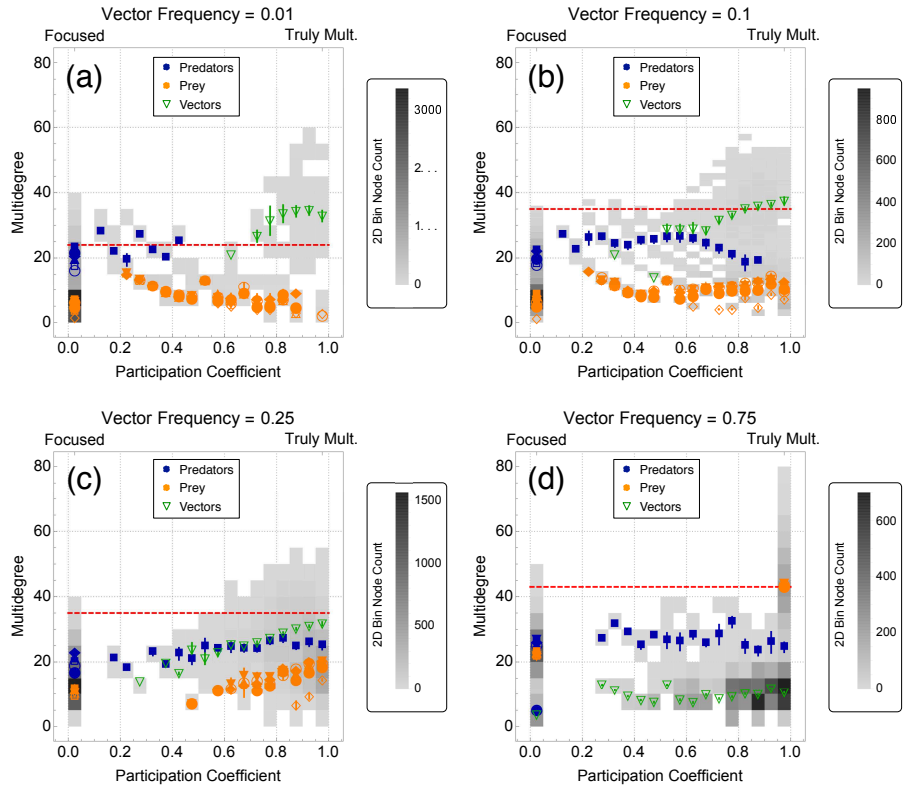}
%
\caption{Cartographies as 2D histograms for the SEMN model for vector frequency $f_v=0.01$ (a), $f_v=0.1$ (b), $f_v=0.25$ (c), and $f_v=0.75$ (d). The 10,000 multiplex nodes are binned in 2D bins, according to their coordinates in the cartography. Bins are colour-coded according to the number of points falling within them: more coloured tiles indicate a higher number of nodes. Coloured dots identify individual species: predators (blue), prey (orange) and vectors (green). Nodes falling above the horizontal red line have degrees above the 95th percentile in the multidegree distribution and they are therefore considered hubs. Error bars represent standard error of the mean.}  
\label{fig7} 
\end{figure}

The SEMN model considers empirical interaction matrices $T_{eco}$ and $V_{eco}$ compared to SMN. Notice that the in $V_{eco}$ the vector contaminates only 7 of the 20 species in the ecosystem, while in SMN it is allowed to contaminate all the other 2 species. In this section we relate the empirical ecological structure to the results for SEMN.
The cartographies reported in Figure \ref{fig7} (a-d) represent snapshots of the spatial ecological multiplex network with increasing frequencies of vectors. In all the cartographies there is one predator species that displays a wide variation in the participation coefficient, while the participation coefficients of the other predator species populations is zero. This is because, differently from SMN, the SEMN model has one predator species that can be contaminated by vectorial transmission (see $V_{eco}$ in the Supporting Information), while the other predator species populations have links only on the trophic layer. When vector populations are rare ($f_v=0.01$), predator and prey populations occupy the same regions of the cartography, as in the SMN model, see Figure \ref{fig7} (a) and (b). A similar case occurs with prey populations, since only half of them have connections on the vectorial layer (see $V_{eco}$ in the Supporting Information). Analogously to the SMN model, increasing the frequency of vectors leads to scenarios where some predator and prey populations display a wide range of participation coefficients. However, at both  $f_v=0.1$ and $f_v=0.25$ predator populations have a higher multidegree than prey populations. This occurs because predators receive more connections than prey in the trophic layer. Therefore, for values as low as $f_v=0.1$ the species types show varied and distinct patterns in the cartography. At $f_v=0.25$, prey populations show an increased participation in the multiplex network as a sign of increased connectivity in the vectorial layer (Figure~\ref{fig7} (c)). When vector populations are highly frequent in the system, $f_v=0.75$, the cartography reveals some extreme patterns: prey species populations that interact with vectors on the vectorial layer display participation coefficient close to 1 while the other prey species show focused interactions  (Figure~\ref{fig7} (d)). This same pattern was observed between predator species  populations that interact with vectors and the predator populations that do not when  $f_v=0.75$ (Figure~\ref{fig7} (d)). This was not observed in the SMN model. 

\begin{figure}
\centering
\includegraphics[width=0.55\paperwidth]{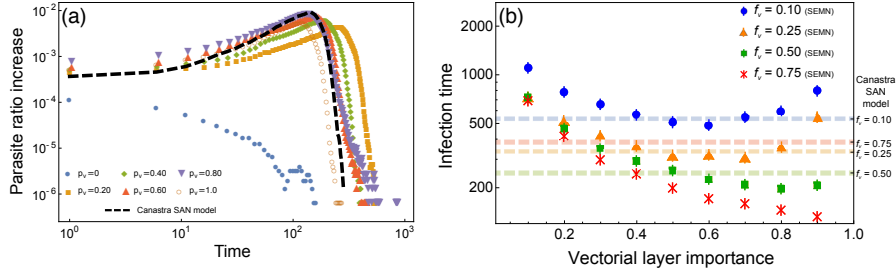}
%
\caption{(a): global parasite ratio increase over time for $f_v=0.75$ for SEMN model and different vectorial layer importance. A qualitatively similar behaviour was observed also for other vector frequencies. (b): global infection time versus vectorial layer importance $p_v$ for different vector frequencies in the SEMN model. Dotted lines represent infection time in the SAN model applied to Canastra empirical data for different vector frequencies. Results in both plots are averages of 100 repetitions.
}  
\label{fig8} 
\end{figure}

As reported in Figure \ref{fig8} (b), the time required to infect almost all the populations in SEMN is minimised when there is a high frequency of vectors in the environment and a high importance of vectorial layer for parasite transmission. Infection times decrease monotonically when  $f_v=0.5$ and $0.75$. However,  at vector frequencies $f_v=0.1$ and $0.25$ parasite spreading is optimised when the vectorial layer importance $p_v$ is around 0.6 (\ref{fig8} (b)), that is, when vectorial and trophic transmission mechanism have similar importance. Therefore, vectorial and trophic transmission mechanism have an additive effect for parasite spreading only when  $f_v<0.5$. Comparing the results against a spatial aggregate network model using the Canastra matrices (Canastra SAN model) reveals how the multiplex structure can change dramatically the infection time. For instance, when $f_v=0.1$, the infection time of the Canastra SAN model is halved compared to the SEMN one for $p_V=0.1$, see also the dashed lines in Figure~\ref{fig8} (b). The multiplex structure not always increases the speed of parasite spreading and the multiple dynamics that resulted from the interplay of vectorial layer importance and community composition justifies the value of investigating different transmission routes via multiplexity.
Despite the higher connectivity of the trophic layer in the SEMN model, parasite ratio increases behave similarly to the SMN model (\ref{fig8} (a)). The parasite spreading propagates much slower on the trophic layer alone than on the full multiplex structure, see the $p_V=0$ trajectory. Again, considering also contaminative interactions provides qualitatively different dynamics of parasite ratio increases than considering trophic interactions only (\ref{fig8} (a)). However, the dynamics of parasite ratio increases in time for the SEMN model are qualitatively similar to the SAN model relative to $p_V>0$. Increasing the vectorial layer importance accelerates the parasite spreading even though no monotonous relationship is evident from the plots. For $p_v>0$ the slow-down phase following the increase peaks does not behave independently of $p_v$ (\ref{fig8} (a)). Therefore, these peaks cannot be considered good proxies of the infection times in the SEMN model. When the spreading deceleration occurs in different time windows, it sums up differently to the peak times, thus establishing global infection times that are not straightforwardly related to the peak times. For instance, the peak for $p_v=0.8$ is reached sooner for the $p_v=0.6$ but the deceleration phase takes longer for $p_v=0.8$ than for the $p_v=0.6$ and $p_v=0.8$ has a higher global infection time compared to $p_v=0.6$. 

The SEMN model also displayed a phase transition in the emergence of a global epidemic, similarly to what happened for the SMN model. However, the different topology of trophic and vectorial layers brought to a slight increase in the critical vector frequency value, from $f_v=0.02$ (SMN) to $f_v=0.04$ (SEMN). 

\subsubsection{Immunisation scenarios in the SEMN model}

\begin{figure}
\centering
\includegraphics[width=0.55\paperwidth]{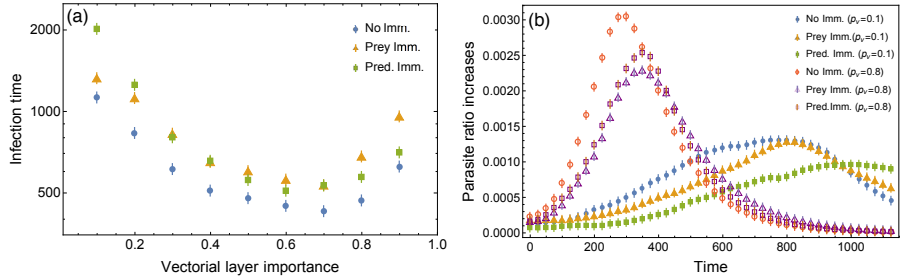}
%
\caption{(a): global infection time versus vectorial layer importance $p_v$ for different immunisation experiments with $f_v=0.75$ in the SEMN model. The no immunisation scenario means that no immunised populations are present in the system while other dot types represent scenarios in which only prey or predators are immunised, respectively. For immunisation scenarios the same number of populations has been immunised. (b): parasite ratio increase of infected nodes over time for the SEMN model for different immunisation scenarios with $f_v=0.75$. Error bars are computed over 10 independent experiments. Differently from the behaviour we observe in the SMN model, immunising prey is less effective than immunising predators in slowing down the disease spread for small $p_v$ values. The opposite scenario happens when $p_v>0.2$ where immunising prey is more effective than immunising predators, as shown in panel (b) comparing $p_v=0.1$ and $p_v=0.8$ immunising scenarios. }  
\label{fig9} 
\end{figure}

Unlike the SMN model, the SEMN model has predator and prey populations exhibiting different cartography patterns only at low vector frequencies. Therefore, we investigated immunisation scenarios at $f_v=0.1$ and $f_v=0.25$. The results for $f_v=0.1$ are reported in Figure \ref{fig9} and are analogous to the $f_v=0.25$ case (plots not shown for brevity). 

Both the SMN and the SEMN models are spatially embedded, but SEMN has a higher number of species with interaction patterns based on empirical data. In SEMN, immunising prey over predator populations does not always hamper more the parasite spreading, as it happened in the SMN model. From the cartography in Figure \ref{fig7} (a) one would expect predator populations to play a pivotal role in spreading the parasite, given their higher multidegree compared to prey populations, on average.  However, in the same cartography 6 out of 12 prey species display a higher average participation coefficient compared to 6 out of 7 predator species (with participation coefficient equal to zero). Hence, from the cartography both predator and prey populations could play a central role in promoting the parasite spreading: predators are hubs while prey can spread the infection across both layers. In contrast to the SMN model, it is not possible to make predictions based on the cartography alone. 
Our immunisation simulations reveal the presence of two scenarios: when the parasite spreads mainly across the trophic layer ($p_v < 0.3$), then immunising the same number $\phi=346$ of predator over prey populations significantly increases the infection times, (\ref{fig9} (a)), and slows down parasite diffusion (Figure \ref{fig9} (b)). This finding relates to the SEMN cartography: predator populations have a high multigree because they are hubs in the trophic layer (here $f_v=0.1$) and hence promote the parasite spreading through trophic interactions. However, when the vectorial layer importance $p_v$ increases above $0.3$, then immunising predator or prey populations does not make noticeable difference. When $p_v>0.7$ and the parasite spreads mainly through contaminative interactions the most effective immunisation strategy becomes immunising prey populations, since vectors contaminate mostly prey populations in the SEMN model (Figure \ref{fig9}). Again, this is compatible with the patterns in the multiplex cartography: when $p_v$ is higher, the multiplex structure becomes predominant and the species populations that have higher participation coefficients, such as prey, can promote the infection spread.

\section{Discussion}

It is only recently that network scientists started addressing the multiplex structure of real-world systems such as ecological and epidemiological systems \cite{kivela2014multilayer,dedomenico2013mathematical,boccaletti2014structure,kefi2015ecological}.  Multi-layer networks were used in ecological systems to approach different interaction types \cite{melian2009diversity, fontaine2011ecological} and levels of organisation \cite{belgrano2005aquatic, scotti2013social,barter2016spots}. More in particular, multiplex networks were used for the first time in \cite{kefi2015multiplex}, in order to consider trophic and non-trophic interactions together in a Chilean ecosystem. In epidemiological systems multi-layer networks were used to describe parasite spreading with Susceptible-Infected-Susceptible dynamics \cite{saumell2012epidemic, sahneh2013effect, granell2013dynamical, sanz2014dynamics,lima2015disease}, susceptible-infected-recovered dynamics \cite{dickison2012epidemics, marceau2011modeling, Buono2014}, and multiple types of interactions between random layers \cite{zhao2014multiple, Cozzo2013, salehi2014diffusion}. The modelling of multi-host parasites that are transmitted through multiple mechanisms in the ecosystem can be improved by applying the framework of multiplex networks. We used the multiplex approach to study both a simple predator-prey-vector system as a reference case, and an empirical data from host communities of \textit{T. cruzi} in natural habitat (Canastra). Compared to their aggregate counterparts, both our multiplex network models displayed a richer phenomenology in terms of infection dynamics. Our three-species-system (SMN) as well as our empirical-based model (SEMN) showed that the epidemiological importance of vectors, hosts and parasites might be mapped on the multiplex cartography. Considering the node and link heterogeneity in a spatial context allowed for us to identify percolation thresholds for parasite spreading according to vector frequency. This is particularly interesting because the susceptible-infected dynamics in homogeneous hosts always leads to epidemic waves (in other words, when nodes are not spatially embedded there is no percolation threshold). In addition, we found that multiplex cartography had important implications in parasite spreading dynamics and that parasite transmission depends on: (i) the relative importance of the distinct transmission mechanisms, (ii) the role species play on the overall multiplex structure and (iii) the species relative frequencies in the system. 

There is a strong debate in ecology on whether biodiversity reduces or not the risk of infection in host communities \cite{johnson2013host,keesing2006effects, wood2014does}. In general, the effect of host diversity on parasite transmission depends on the ecological characteristics of hosts and on the mechanism of transmission \cite{wood2014does}. The spatial multiplex modelling framework that we propose in this study could be applied to address questions related to the role of multiple host community biodiversity on parasite transmission. In fact, we found that the spatial component has a significant impact on the speed of parasite spreading: spatial correlations slowed the speed of parasite spreading when compared to mean-field approximations. Therefore, considering the spatial structure of host communities in order to infer the importance of different host species for parasite transmission is a fundamental next step in future ecological disease studies \cite{craft2011network,kefi2015ecological}. Percolation thresholds are spatially explicit tipping points that indicate the presence, in some regimes, of non-local correlations within a given system \cite{davis2008abundance}. For instance, if a network is not strongly connected, then the parasite will not be transmitted to the whole system. In our model the connectivity of the multiplex network was crucially affected by the frequency of different species. For very small frequency of vectors $f_v$, our model showed a percolation threshold in both the SMN and the Canastra SEMN model. The presence of such phase transition in the infection rate in an SI dynamics for a non-zero value of  $f_v$ is mainly related to (i) the spatial structure and to (ii) directed trophic interactions in the multiplex network. In the SMN model the parasite can percolate through the whole system only if $f_v>0.02$, while in the Canastra SEMN model the critical vector frequency was found to be around $f_v=0.04$. No phase transition for $f_v>0$ was found in the RAN model, where nodes are not spatially embedded. We conjecture that the increase in the percolation threshold from the SMN to the SEMN models might be due to a higher diversity of potential hosts: with more species available there is an increased chance that vectors will interact with animals that do not become infected with the parasite. Interestingly, our theoretically computed frequencies agree with previous findings that even a small frequency of vectors in the ecosystem is sufficient to maintain Chagas disease in a human population \cite{reithinger2009eliminating}. 

Multiplex cartography \cite{battiston2014structural} considers both the relative frequency of each species and the interactions they have in both the trophic and the vectorial layers. Comparisons with aggregated networks revealed that considering trophic and vectorial transmission routes together can change dramatically the parasite spreading dynamics, depending on the relative frequency of vectors in the ecosystem. More in detail, the parasite spreading dynamics depends on the interplay between community species composition and the relative importance of the transmission mechanisms. In fact, when there is homogeneity in species composition (i.e. when the relative frequency of vectors $f_v\sim 0.5$), the lowest infection time is registered when the parasite spreads on both layers at the same time (i.e. for intermediate values of $p_v$) in both the SMN and the SEMN models. Therefore, our theoretical network models indicate that vectorial and trophic mechanisms of transmission can be additive in sustaining the spread of multi-host parasites such as \textit{T. cruzi}, further agreeing with previous studies \cite{kribs2006vector}. In random multiplex networks \cite{saumell2012epidemic}  the epidemic process also depends on the strength and nature of the coupling between the layers. In our case the vectorial layer importance $p_v$ can be thought of as an implicit coupling between the layers, quantifying how much the vectorial layer is more important than the trophic layer in spreading the parasite. Previous investigation \cite{saumell2012epidemic,dedomenico2013mathematical,kivela2014multilayer,boccaletti2014structure} showed that epidemic dynamics on a multiplex structure can be fundamentally different from the same dynamics on each multiplex layer considered as separate. Our results indicate that multiple mechanisms may speed up parasite spreading. The multi-layered transmission, which is observed in many parasites with complex life cycles and multiple mechanisms of infection, seems to be a very efficient strategy for spreading in communities of multiple hosts. 

In vector-borne diseases, densities of hosts and vectors as well as the ratio of their densities, have strong implications for parasite transmission \cite{ross1911case, velascohernandez1994model, kribs2010alternative,pelosse2012role}. The SMN model shows that higher vector frequencies make the vectorial layer faster in spreading the parasite from vectors to predator and prey populations. This relationship explains why infection times decrease monotonically with increased importance of the vectorial layer. On the other hand, if the vector frequency is low and the parasite spreads only on the trophic layer, it becomes increasingly difficult to infect more populations over time. In this situation the fastest global infection is achieved when both mechanisms of transmission are likewise selected for parasite spreading (there is a minimum in the infection time around $p_v=0.6$). Moreover, in the Canastra SEMN model, we observe an analogous minimum even with higher vector frequencies. This suggests that global infection time is minimised when both mechanisms of transmission have similar importance in more complex ecological scenarios. Notice that considering both the transmission mechanisms but with one layer much more important than the other (e.g. $p_v = 0.1$) can lead to drastic increases in the infection time. The evolution and maintenance of mutually important multiple routes of transmission may be selected in parasites that infect a high number of host species. 

Furthermore, using the multiplex cartography we predict that the relative importance of each mechanism for parasite spreading depends on the host community composition and relative frequency of species. We find that species structural patterns, encapsulated within the multiplex cartography, are a valuable measure to evaluate the importance of each species for parasite spreading. These findings are confirmed by the immunisation simulations. For instance, in the SMN model, a higher frequency of vectors ($f_v>0.5$) increases prey populations connectivity and therefore their participation in the multiplex topology. We find different results when considering a more realistic ecological scenario. In the SEMN model, predator populations dominated the multiplex topology because of their higher connectivity and higher average multidegree. Immunising prey populations in the reference SMN model dramatically increases global infection time and the rate of disease spreading in the populations. However, in the SMN model immunising prey over predators results in different infection times only when these species occupy distinct regions in the multiplex cartography. This result points to the meaningfulness of the network cartography for understanding the parasite spreading dynamics. In fact, the multiplex cartography shows that prey participate more and have higher degree in the three-species multiplex network and thus could be a better target for immunisation. The immunisation simulations confirm this: immunising prey populations hampers the parasite spreading with respect to immunising the same number of predator populations. In the Canastra SEMN model, predators are the species type that attain most of their connections in the multiplex network and thus have a higher importance in the cartography. This pattern suggests that the predators are acting as a sink for the parasite and can thus reduce the overall parasite transmission in the SEMN model. This is mainly due to the fact that predators are hubs in the trophic layer and hence show a higher multidegree in the cartography. When the parasite spreads mainly in the trophic layer ($p_v<0.3$) the immunisation experiments indicate that immunising predators hampers the disease more compared to immunising prey. This is in agreement with empirical studies pointing out the potential importance of predators as parasite bio-accumulators \cite{rocha2013carnivores, jansen2015multiple}. However, prey also display a slightly higher average participation in the Canastra cartography and hence could also play a central role in spreading the parasite. In fact, when the vectorial layer importance $p_v$ is above 0.7, immunising prey populations becomes the most effective immunisation strategy. This is because vectors contaminate mostly prey in the Canastra multiplex network. Again, the roles played by each species in the multiplex cartography depended on the frequency of vectors and is related to their importance for parasite spreading.

It has to be underlined that the main aim of our multiplex model is not to provide a realistic mechanism for the spreading dynamics of \textit{T. cruzi} in wild hosts. Instead, our approach aims at providing a comprehensive framework for investigating the spreading of multi-host parasites across different transmission mechanisms. Additional information should be taken into account if one would want to study the dynamics of \textit{T. cruzi} in wild hosts and Chagas disease epidemiology.  For instance, it is known that the stercorarian transmission results in a much higher probability  of parasite transmission from host to vector than from vector to host \cite{rabinovich1990probability}. More realistic models should include these differences via different contact rates on different layers. In addition, host physiological and ecological characteristics influence their probability to transmit \textit{T. cruzi}. A higher proportion of insects in host diets increase host probability of infection \cite{roellig2009genetically, rabinovich2011ecological, rocha2013carnivores}. Finally, host species that share ecological habitat with vector species are more likely to be exposed to the infection \cite{jansen2015multiple}. 
Many zoonoses, which are infections naturally transmitted between vertebrate animals and humans, may have multiple hosts and mechanisms of transmission. Examples of zoonoses transmitted to humans by  arthropod vectors include Malaria, Leishmaniasis, Chagas disease, West Nile virus, plague and Lyme disease \cite{schmidt2001biodiversity}. The multiplex framework presented here could improve our understanding of the epidemiology and evolution of these parasites and help us elaborate more efficient control strategies for reducing disease incidence in humans. For instance, different or additional layers could be included within our multiplex framework to make the model more realistic, such as direct transmission mechanism or the network of human interactions with its socio-ecological characteristics. Outside of the ecological perspective, our spatial multiplex network model could be applied to modelling systems made of spatially embedded interacting agents where instead of parasite infection there is a given information spreading process.

\subsection*{Author Contributions}

A.A., C.S.A., S.S. and M.S. conceived and designed the study, A.A. and A.G. wrote the code, A.A., A.G. and M.S. performed experiments, C.S.A. analysed the empirical data and M.S. analysed the simulation data, A.G. and M.S. conceived the analytical part, C.S.A., S.S. and M.S. discussed the results. All authors wrote the paper and gave final approval for publication.

\subsection*{Supplementary Material}

The vectorial and trophic matrices for the Canastra SERN model have been uploaded as supplementary material.

\subsection*{Acknowledgements}

We thank Hans Heesterbeek, Mason A. Porter, Darko Stefanovic, Jennifer Dunne, Markus Brede, the anonymous manuscript reviewers and the Santa Fe summer school scientific committee for their feedback on the early version of the manuscript. This work was supported by Conselho Nacional de Desenvolvimento Cient\'\i fico e Tecnol\'ogico [CNPq/Brazil to C.S.A.], the NSF [\#1028238 and \#1028120 to A.G.], the Complexity programme of The Netherlands Organisation for Scientific Research (NWO to S.S.), the Doctoral Training Centre at the University of Southampton and the EPSRC [EP/G03690X/1 to M.S.], the Swiss National Science Foundation [200020-143224, CR13I1-138032 and P2LAP1-161864 to A.A.] and by the Rectors' Conference of the Swiss Universities [26058983 to A.A.].

\nocite{*}

%

\end{document}